\documentclass[fleqn,usenatbib]{mnras}

\usepackage{newtxtext,newtxmath}
\usepackage{xcolor}
\usepackage[normalem]{ulem}

\usepackage{orcidlink}
\usepackage[normalem]{ulem}

\usepackage[T1]{fontenc}

\DeclareRobustCommand{\VAN}[3]{#2}
\let\VANthebibliography\thebibliography
\def\thebibliography{\DeclareRobustCommand{\VAN}[3]{##3}\VANthebibliography}

\usepackage{graphicx}	%
\usepackage{float}
\usepackage{physics}
\usepackage{amsmath}	

\title[Accelerated binary black holes in globular clusters]{Accelerated binary black holes in globular clusters: forecasts and detectability in the era of  space-based gravitational-wave detectors}

\author[A. Tiwari et al.]{Avinash Tiwari \orcidlink{0000-0001-7197-8899},$^{1}$\thanks{E-mail: avinash.tiwari@iucaa.in} 
Aditya Vijaykumar \orcidlink{0000-0002-4103-0666},$^{2}$ 
Shasvath J. Kapadia \orcidlink{0000-0001-5318-1253},$^{1}$
Giacomo Fragione \orcidlink{0000-0002-7330-027X}$^{3,4}$ and
\newauthor
Sourav Chatterjee \orcidlink{0000-0002-3680-2684}$^{5}$
\\
$^{1}$Inter University Centre for Astronomy and Astrophysics, Post Bag 4, Ganeshkhind, Pune - 411007, India\\
$^{2}$International Centre for Theoretical Sciences, Tata Institute of Fundamental Research, Bangalore 560089, India\\
$^{3}$Department of Physics $\&$ Astronomy, Northwestern University, Evanston, IL 60208, USA
\\
$^{4}$Center for Interdisciplinary Exploration $\&$ Research in Astrophysics (CIERA), Northwestern University, Evanston, IL 60208, USA\\
$^{5}$Tata Institute of Fundamental Research, Homi Bhaba Road, Navy Nagar, Colaba, Mumbai 400005, India}

\pubyear{2023}

\begin{document}
\label{firstpage}
\pagerange{\pageref{firstpage}--\pageref{lastpage}}
\maketitle

\begin{abstract}
The motion of the center of mass of a coalescing binary black hole (BBH) in a gravitational potential, imprints a line-of-sight acceleration (LOSA) onto the emitted gravitational wave (GW) signal. The acceleration could be sufficiently large in dense stellar environments, such as globular clusters (GCs), to be detectable with next-generation space-based detectors. In this work, we use outputs of the \textsc{cluster monte carlo (cmc)} simulations of dense star clusters to forecast the distribution of detectable LOSAs in DECIGO and LISA eras.
We study the effect of cluster properties---metallicity, virial and galactocentric radii---on the distribution of detectable accelerations, account for cosmologically-motivated distributions of cluster formation times, masses, and metallicities, and also incorporate the delay time between the formation of BBHs and their merger in our analysis.
We find that larger metallicities provide a larger fraction of detectable accelerations by virtue of a greater abundance of relatively lighter BBHs, which allow a higher number of GW cycles in the detectable frequency band. Conversely, smaller metallicities result in fewer detections, most of which come from relatively more massive BBHs with fewer cycles but larger LOSAs. We similarly find correlations between the virial radii of the clusters and the fractions of detectable accelerations. Our work, therefore, provides an important science case for space-based GW detectors in the context of probing GC properties via the detection of LOSAs of merging BBHs.
\end{abstract}




\section{Introduction}

The formation, evolution, and merger environments of binary black holes (BBH) are subjects of many active research efforts (see eg. \citealt{Mapelli:2021taw} for a review). 
The prevalent expectation is that the majority of the BBHs detected by the LIGO-Virgo-KAGRA network \citep{aligo, avirgo, kagra} likely formed either through isolated evolution in the galactic field or through many-body interactions in dense dynamical environments. Isolated evolution could proceed {mainly} via a common envelope phase \citep{belczynski2016, giacobbo2018, kruckow2018}\footnote{Note however that some works \citep{2019MNRAS.490.3740N, 2017MNRAS.471.4256V,2021ApJ...922..110G, 2022ApJ...940..184V} are finding that a majority of binaries do not require a common envelope phase and could form and evolve just via stable mass transfer.}, or via chemically homogeneous evolution \citep{demink2016, marchant2016}. Dynamical environments could include globular clusters \citep[GCs;][]{askar2017, banerjee2010, banerjee2018, Chatterjee2017, Chatterjee2017a, fragione2018, rodriguez2018, dicarlo2020, kremer2020, mapelli2021, trani2021, FragioneRasio2023}, nuclear star clusters \citep{antonini2012, petrovich2017, grishin2018, hoang2018, fragione2020}, and disks of active galactic nuclei \citep[AGN;][]{bartos2017, secunda2019, li2021, ford2022}, among others.

The $\sim 90$ BBH detections reported by the LIGO-Virgo-KAGRA collaboration \citep{LIGOScientific:2021djp} have started to shed some light on their origin \citep{gwtc3-rp}.
However, making precise inferences on formation channels from data needs to take into consideration two factors. The first is that detected binaries could be coming from a combination of the aforementioned formation channels. 
Indeed, the data suggest that multiple formation sub-channels even within isolated evolution contribute to this spectrum, although the extent of these contributions from different channels is unknown and not straightforward to constrain, in part because of the systematics associated with the population synthesis simulations \citep{WongBreivik2021,zevin2021}.
The second is that in general, the shape of the GW waveform itself contains no definite signatures that can conclusively ascertain the provenance of the binary on a single-event basis

{In the case of BBH mergers assembled dynamically,} the binaries move on orbits determined by the star cluster gravitational potential. As this motion could leave an imprint on the GW signal {in the form of a Doppler shift}, its detection would contribute to our ability to identify the binary formation channel\footnote{{There exist a number of other studies too, in the literature, done on the
identification of the formation channel of BBHs by looking at imprints of the formation channel on the GW signal. 
A few of them are \cite{2019MNRAS.488.5665W}, \cite{2020PhRvD.101h3031D}, and \cite{2021PhRvD.104j3011Y} that consider the modulations to the GW waveform due to at least one among Doppler shift, repeated gravitational lensing, and de Sitter precession.}}.  {However,} a binary orbit at constant velocity would produce a constant Doppler shift in the GW waveform, degenerate with the mass of the binary. {On the other hand}, accelerated motion (with a non-zero component of the acceleration along the observer's line-of-sight) could modulate the signal {and, therefore, be detectable} \citep{yunes2011, bonvin2017, tamanini2020, vijaykumar2023}. Constraints on this line-of-sight acceleration inferred directly from the GW signal could hence carry information on {the environment} in which the binary merged \citep{vijaykumar2023}. 

GCs are among the dense stellar environments expected to efficiently assemble BBH mergers. They are stable, spherically symmetric, gravitationally bound collections of $\sim 10^4 - 10^6$ stars with typical sizes of $\sim 1-10$~pc \citep{Harris1996,gratton2019,BaumgardtVasiliev2021}. {BBHs merging in GCs are expected to present an acceleration reminiscent of the environment in which they formed}. Thus, detecting signatures of (time-varying) Doppler shift could not only {point towards} identifying different formation environments, but could also provide {crucial information} about masses, {density} profiles, metallicities, and ages of GCs.

In this work, we calculate accelerations of BBHs in GCs, as a function of the cluster properties, using the catalogue pertaining to the large-scale \textsc{cluster monte carlo (cmc)} \citep{cmc-paper} simulation. We extract and determine the accelerations of all the BBH binaries that merge within a Hubble time from the \textsc{cmc} catalogue, and employ a GW Fisher analysis\footnote{{As per the Cramer-Rao bound, the Fisher analysis gives the most-optimistic values of the uncertainties in the parameters. The results presented in this work should thus be thought of as best-case estimates.}} \citep{cutlerflanagan} to estimate whether such accelerations can be sufficiently well constrained with the proposed DECIGO \citep{sato2017} and LISA \citep{danzmann2003} space-based detectors.  We construct distributions of accelerations as a function of GC properties, with appropriately chosen detectability, metallicity, and cluster-mass weights. We then study the imprint of GC properties on these distributions. 

The rest of the paper is organised as follows. Section~\ref{sec:methods} describes the \textsc{cmc} catalog models and outlines the prescription we use to construct distributions of BBH accelerations in GCs. Section~\ref{sec:results} presents the results and, in particular, the imprint of GC properties on the distribution of accelerated BBHs. Section~\ref{sec:conclusions} summarizes the paper, discusses this work in the context of other GW probes of GCs, and suggests the scope for future work. In the entirety of the paper, we assume the standard cosmological model with parameters fixed to the Planck 2018 values \citep{planck18}.

\section{Method}\label{sec:methods}

\subsection{The \textsc{cmc} models}
The \textsc{cmc} catalogue comprises $144$ simulations of GCs. It uses a H\'enon type Monte Carlo algorithm which enables a long-term evolution of the GC \citep{henon1971a, henon1971b, joshi2000, joshi2001, fregeau2003, fregeau2007, chatterjee2010, chatterjee2013, pattabiraman2013, rodriguez2015}, assuming a set of initial conditions. Details of the \textsc{cmc} simulation can be found in \citealt{cmc-paper, cmc-url}. Here, we briefly summarize some of the most important features of the models.

Four {different} initial {cluster} properties {describe} the \textsc{cmc} catalogue grid. These properties are: the total number of single stars and binaries in the cluster ($N = 2 \times 10^5, 4 \times 10^5, 8 \times 10^5, 1.6 \times 10^6$), the initial virial radius of the cluster ($r_v/\mathrm{pc} = 0.5, 1, 2, 4$), the galactocentric radius of the cluster ($r_g/\mathrm{kpc} = 2, 8, 20$), and initial metallicity of the cluster ($Z = 2 \times 10^{-2}, 2 \times 10^{-3}, 2 \times 10^{-4}$). Each combination of these parameters corresponds to one \textsc{cmc} simulation and the outputs of all the $144$ simulations are catalogued in \citep{cmc-url}.

A number of fixed initial conditions are assumed for the whole set of simulations. The initial cluster potential is assumed to follow a King profile \citep{king1962}, with concentration parameter $W_0 = 5$.  The stellar masses are drawn from a Kroupa initial mass function \citep[IMF;][]{kroupa2001}, assuming a mass range of $0.08 \mathrm{M}_{\odot}- 150 \mathrm{M}_{\odot}$ and the stellar binary fraction is set to $f_\mathrm{b} = 5\%$. 
For binaries, the primary component is drawn from a Kroupa IMF, while the secondary component is chosen by drawing from a uniform distribution of mass ratios $q \in [0.1, 1]$. The initial orbital period {of binaries} is drawn from a log-uniform distribution, with a lower limit on the separation set such that this separation ($d$) does not fall below five times the sum of the stellar radii of the binary ($d \geq 5(R_1 + R_2)$), and an upper limit set by the hard/soft boundary. Each simulation is evolved across $14$ Gyr or until the GC undergoes tidal disruption (see eg.~\citealt{2003gmbp.book.....H}) or collisional runaway (see eg.~\citealt{PortegiesZwart:2002iks}).

A number of physical processes have been incorporated into the \textsc{cmc} simulations. These include stellar and binary evolution, neutron star formation, black hole (BH) formation, modeling of strong encounters, two-body relaxation, three-body binary formation, implementation of galactic tides, and stellar collisions. We refer the reader to \citealt{cmc-paper} for details on all these processes; we briefly summarize the prescriptions used for BH formation below.

BHs are modeled to form via standard iron core-collapse supernovae (CCSNe) using the ``rapid model'' for stellar remnants \citep{fryer2012}. The CCSNe impart natal kicks to the BH, with mass fallback decreasing the magnitude of the kick. The kick velocities of neutron stars, $V_\mathrm{NS}$, are assumed to be described by a Maxwellian distribution with a dispersion set to $\sigma = 265 \mathrm{km/s}$. {For BHs,} the kick magnitudes are then modulated as a function of the fallback mass fraction $f_\mathrm{b}$ such that $V_\mathrm{BH} = (1 - f_\mathrm{b})V_\mathrm{NS}$, where this fraction pertains to the percentage of stellar envelope mass that falls back onto the collapsed core. Additionally, pulsational pair-instability~\citep{pisn} is implemented, which results in the mapping of stars with helium core masses in the range $45 - 65 \mathrm{M}_{\odot}$ to BHs of masses in the vicinity of $40 \mathrm{M}_{\odot}$ \citep{Woosley2017}, producing an excess in that region of the BH mass spectrum. Stars with helium cores in excess of $65 \mathrm{M}_{\odot}$ are modeled to produce no remnants at all \citep{HegerFryer2003}. 

\begin{figure*}
    \centering
    \includegraphics[width=0.9\linewidth]{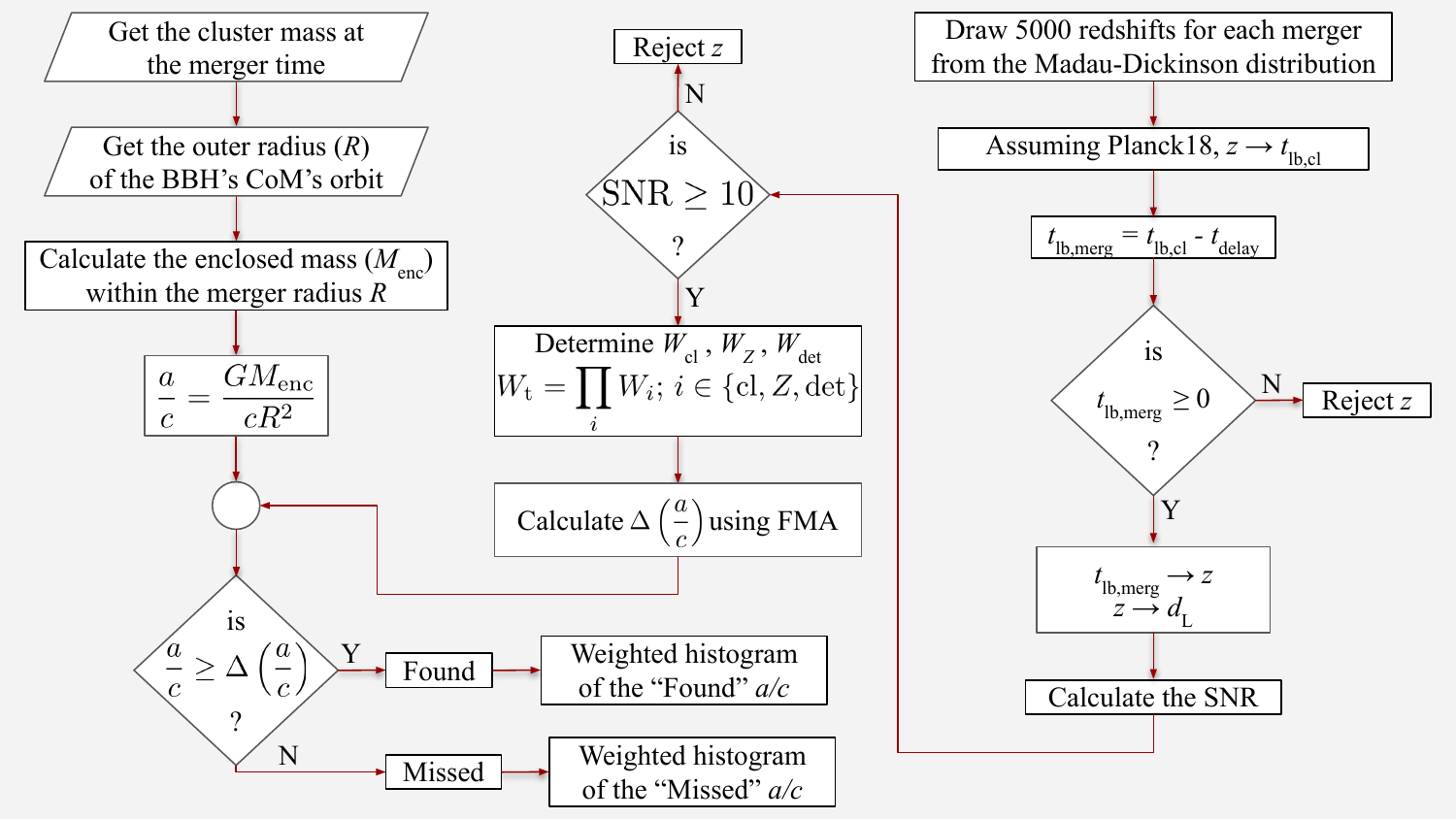}
    \caption{A flowchart summary of the prescription to extract accelerations from the \textsc{cmc} simulations, and construct weighted distributions of identifiable (found) and non-identifiable (missed) BBH accelerations ($a/c$). A standard FRW metric is assumed, with cosmological parameters taken from ``Planck18'' \citep{planck18}. CoM is the ``centre of mass'', $t_\mathrm{lb, cl}$ is the lookback time of the GC at formation, $t_\mathrm{delay}$ is the time to coalescence from the formation of the binary, which is assumed to coincide with the formation time of the cluster, and $t_{\rm lb, merg}$ is the lookback time at merger.}
    \label{fig:flowchart}
\end{figure*}

\subsection{Extracting accelerations from the \textsc{cmc} catalog}

We describe below the prescription used to evaluate the accelerations of merging BBHs in GCs. A flowchart summary of the prescription we use below is provided in Figure~\ref{fig:flowchart}.

\begin{enumerate}
    \item For each merger in the \textsc{cmc} catalog, we determine the mass of the cluster $M_{\mathrm{enc}}$ enclosed within a radius $R$, where $R$ is the distance of the BBH from the centre of the cluster when it merges. 
    \item The acceleration of the center of mass of the BBH divided by the speed of light $a/c$, is then evaluated as \citep{BinneyTremaine,BovyBook}:
    \begin{equation}
        a/c = GM_{\mathrm{enc}}/cR^2
    \end{equation}
    \item For each BBH and corresponding acceleration, $n = 5000$ redshift samples are drawn following the cosmic star-formation rate density (SFRD) as given in the Madau-Fragos prescription~\citep{madaufragos2017, MD}:
    \begin{equation}
        p(z) \propto \frac{(1 + z)^{2.6}}{1.0 + [(1.0 + z)/3.2]^{6.2}}
    \end{equation}
    These samples correspond to cluster-formation redshifts. In essence, we assume that the history of cluster formation follows that of stars.
    \item To evaluate the merger epochs, we first convert the cluster-formation redshifts to lookback time  $t_{\mathrm{lb}, \mathrm{cl}}$. Then, using the time-delay values from the simulation, $t_{\mathrm{bbh}, \mathrm{delay}}$, the lookback time of the BBH at merger is calculated as $t_{\mathrm{lb}, \mathrm{bbh}} = t_{\mathrm{lb}, \mathrm{cl}} - t_{\mathrm{bbh}, \mathrm{delay}}$. If positive, this lookback time is now converted back to a redshift at the merger. Otherwise, the sample is rejected, since it implies that the BBH will not merge within the age of the universe.
    \item Converting the redshift at merger to a luminosity distance at merger, and using the intrinsic parameters of the BBH provided from the simulation, the signal-to-noise ratio (SNR) $\rho$ in DECIGO and LISA are calculated as:
    \begin{equation}
        \rho^2 = 4\mathrm{Re}\int_{f_{\mathrm{min}}}^{f_{\mathrm{max}}}\frac{|\Tilde{h} (f)|^2}{S_\mathrm{n}(f)}df
    \end{equation}
    where $\Tilde{h}$ is the Fourier transform of the GW waveform as seen by (projected onto) the detector and modulated by $a/c$, $S_\mathrm{n}(f)$ is the detector's sky-averaged noise power spectral density (PSD), and $f_{\mathrm{min}}, f_{\mathrm{max}}$ are frequency limits set by the detector bandwidth
    \footnote{The detector antenna pattern will change appreciably over the inspiral timescale. However, while calculating the SNR, we do not account for the effects of this time-varying detector antenna pattern for computational ease. We do not expect a significant change in the obtained SNR due to this effect.}. 
    To choose the frequency limits for the mergers, assuming an observation time of 4 years, we follow \cite{PhysRevD.71.084025} with $\{f_\mathrm{min}, f_\mathrm{max}\}$ being $\{10^{-2}, 10\}\mathrm{Hz}$ for DECIGO and $\{10^{-4}, 1\}\mathrm{Hz}$ for LISA. To model $\Tilde{h}$, we use the \textsc{TaylorF2} prescription given by:
    \begin{equation}
        \Tilde{h}(f) = \mathcal{A} f^{-7/6} e^{i (\Psi(f) + \Delta \Psi(f))}
    \end{equation}
    where $\mathcal{A} \propto \mathcal{M}^{5/6}/D_\mathrm{L}$ with $\mathcal{M}, D_\mathrm{L}$ as the chirp mass and luminosity distance of binary respectively, $\Psi (f)$ is given by Eq. (3.18) of \citealt{PhysRevD.80.084043}, and $\Delta \Psi (f)$ is given by Eq. (4) of \citealt{vijaykumar2023}. The BBHs are assumed to have face-on (inner) \textbf{circular} orbits\footnote{{It is worth mentioning that in general, the inner binary is not necessarily circular at sub-Hz frequencies \citep{Breivik_2016}. This is especially true in GCs where a non-trivial fraction of binaries could be eccentric \citep{10.1093/mnras/sty2334, 10.1093/mnras/sty2568, PhysRevLett.120.151101}.}}. 
    \item If $\rho \geq 10$ (8) for DECIGO (LISA), the BBH is considered to be detectable, and the sample is kept. Otherwise, it is rejected. 
    \item Each detected BBH sample is assigned a set of weights: a cluster-mass weight, $W_{\mathrm{cl}}$, and a metallicity weight, $W_{Z}$, to account for the relative cosmological abundance of clusters with different properties. We also assign a detectability weight $W_\mathrm{det}$. These weights are computed following \citealt{fragione2021}. The cluster-mass weight is assigned following the cluster initial mass function as \citep{portegies2010}:
    \begin{equation}\label{eq:cluster-weight}
        W_{\mathrm{cl}} \propto \frac{1}{M_\mathrm{cl}^2}
    \end{equation}
    where $M_\mathrm{cl}$ is the mass of the cluster at formation. The metallicity weight $W_{Z}$ is assigned using lognormal distribution with a $0.5$ dex standard deviation, and redshift-dependent mean given by \citep{madaufragos2017}:
    \begin{equation}
        \log(Z/Z_{\odot}) = 0.153 - 0.074z^{1.34}
    \end{equation}
    The detectability weight is given by:
    \begin{equation}
        W_{\mathrm{det}} = p_{\mathrm{det}}(m_1, m_2, z)\frac{1}{1 + z}\frac{dV_\mathrm{c}}{dz}
    \end{equation}
    where the detection probability $p_{\mathrm{det}}$ is accounted for by setting an SNR threshold and rejecting samples that do not exceed that threshold. As mentioned before, the threshold is 8 for LISA and 10 for DECIGO. To the surviving samples, we then assign a weight determined by the product of the cosmological time dilation piece $1/1+z$ and the differential comoving volume $dV_\mathrm{c}/dz$.
    \item To determine if the acceleration of a sample BBH is constrainable, we resort to a Fisher Matrix Analysis (FMA) which approximates the shape of the GW parameter estimation likelihood to be Gaussian in the source parameters \citep{cutlerflanagan}. From the corresponding covariance matrix, a statistical r.m.s. error $\Delta (a/c)$ is calculated, assuming the Gaussian is centered on true $a/c$. If $a/c < \Delta (a/c)$, $a/c = 0$ is contained within the $68\%$ errorbar, and the BBH's acceleration is said to be ``missed'' ie. the event cannot be confidently identified as accelerating. If $a/c > \Delta (a/c)$, $a/c = 0$ lies outside the $68\%$ errorbar and the BBH's acceleration is said to be ``found''. In other words, the event can be identified as accelerating at 68\% CL. We briefly describe the application of the FMA to the identification of found-missed accelerations in Section \ref{sec:fma}.
    \item We construct various histograms, including histograms of found and missed accelerations, weighted by $W_\mathrm{t}$, the product of the mass, metallicity, and detectability weights \citep{fragione2021}: 
    \begin{equation}
        W_{\mathrm{t}} = W_\mathrm{cl}\times W_{Z} \times W_\mathrm{det}
    \end{equation}
We point out here that the BBH is assumed to be optimally oriented in a way that maximizes the SNR and the magnitude of the LOSA. The latter is also assumed to be unchanging. The inner orbit is assumed to be face-on, while the outer orbit (i.e., the orbit of the BBH's center of mass in the potential of the globular cluster) is assumed to be edge-on. The fractions of found accelerations in this work should therefore be considered as upper limits. 

{To assess the drop in the fraction of measurable accelerations due to a randomized orientation of the outer orbit, as well as due to a more stringent metric for measurability ($2-\sigma, 3-\sigma$ confidence), see Apendix~\ref{app:random_inclination}}.
\end{enumerate}

\subsection{Identifying found-missed BBH accelerations}
\label{sec:fma}

A constant line-of-sight velocity component of the center of mass of a BBH will produce a constant Doppler shift that is degenerate with the mass of the BBH. On the other hand, a BBH with a LOSA will result in a time-varying Doppler shift, which in turn will modulate the GW waveform with respect to one that is not accelerated. At leading order, a deviation $\Delta\Psi(f)$ in the GW phase $\Psi(f)$ is incurred at $-4$ Post Newtonian (PN) order, and is given by \citep{bonvin2017}:
\begin{equation}
    \Delta\Psi(f) = \frac{25}{65536\eta^2}\left(\frac{GM}{c^3}\right)\left(\frac{a}{c}\right)v_f^{-13}
\end{equation}
where $v_f = (\pi GMf/c^3)^{1/3}${, $M$ is the total (detector frame) mass of the binary, and $\eta$ is the symmetric mass ratio}. \citealt{vijaykumar2023} calculated 3.5 PN corrections beyond the leading order to $\Delta \Psi(f)$, and also showed that including these higher-order corrections is necessary for unbiased source property inference. We hence use the full expression of $\Delta \Psi (f)$ from \citealt{vijaykumar2023} to construct our waveform approximant $h(f)$. 

To calculate the r.m.s error $\Delta (a/c)$, the Fisher matrix $\bf \Gamma$ is first constructed as \citep{cutlerflanagan}:

\begin{equation}
	\Gamma_{ij} = \left (\frac{\partial h}{\partial \theta_i} \Big | \frac{\partial h}{\partial \theta_j} \right)
\end{equation}

where $\theta_{i,j}$ are the binary's intrinsic and extrinsic parameters that determine the shape of the GW, and $(|)$ represents a noise-weighted inner product between two GW waveforms $a(f), b(f)$:
\begin{equation}
	(a | b) = 2 \int_{f_\mathrm{min}}^{f_\mathrm{max}} \dfrac{a(f) b^*(f) + a^*(f) b(f)}{S_n(f)} df
\end{equation}
The covariance matrix is then evaluated as $\bf C = \bf \Gamma^{-1}$, and $\Delta(a/c)$ is read-off (and square-rooted) from the corresponding diagonal element in $\bf C$. 

{Intuitively, one would expect that a signal whose phase difference can be tracked for a longer time in-band would provide a better measurement of the acceleration.
Since we choose to model the GW phase in the frequency domain, the tracking of the phase in time is equivalently converted to tracking the phase in frequency and is automatically taken into account by the choice of the bandpass of the detector (i.e. $f_{\min}$, $f_{\rm max}$). Since time in-band $t \approx M^{-5/3} f_{\rm min }^{-8/3}$, this also means that less massive events would have the least measurement uncertainty owing to their long in-band times.}

\section{Results}\label{sec:results}

In this section, we provide weighted distributions of found and missed accelerations, as well as the corresponding fractions with respect to the total number of detected BBHs. We also evaluate the fractions and the distributions as a function of the cluster properties (metallicity, galactocentric radius, and virial radius) in the \textsc{cmc} models. We restrict our attention to DECIGO and LISA detectors, which, by virtue of their sensitivity in the low-frequency regime, are especially suited to detect LOSAs from GCs. 

\subsection{Aggregate distributions of found and missed accelerations in DECIGO}

There are two competing effects that determine if a BBH is detectable and its acceleration is found. {BBHs with heavier masses produce GW signals with larger amplitudes} and are therefore relatively easier to detect, although increasing the mass eventually reduces the detectability {due to a smaller} number of GW cycles in the detector frequency band. On the other hand, {BBHs with} lighter masses produce GWs with a smaller amplitude and are therefore relatively more difficult to detect out to large distances. 

Given that LOSA modulations to the GW are a low-frequency effect, incurring corrections in the GW phase at $-4$PN, longer durations of the in-band inspirals enable stronger constraints on the acceleration, or, equivalently, allow probes of smaller accelerations. {BBHs with} lighter masses spend a longer time in-band relative to {BBHs with} heavier masses, and thus contribute more significantly to the distribution of found accelerations. Moreover, the metallicity and cluster-mass weights also contribute to the fraction of found accelerations, as well as the shape of their distributions (see Appendix~\ref{app:weights}). 

\begin{figure}
    \centering
    \includegraphics{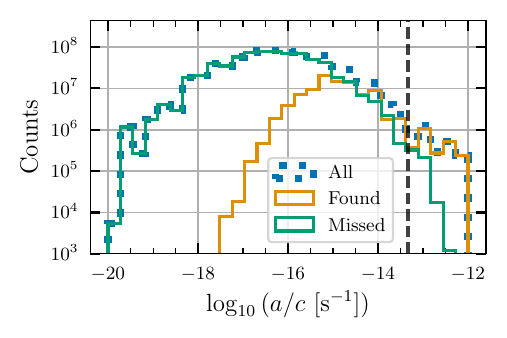}
    \caption{Weighted distributions of detectable, found, and missed accelerations of BBHs in GCs. {The dotted line in the histogram represents an acceleration, of magnitude $4.65 \times 10^{-14} \mathrm{s}^{-1}$, corresponding to a fiducial GC enclosing $10^4 \mathrm{M}_{\odot}$ within a sphere of radius $10^{-2} \mathrm{pc}$}. About $12\%$ of the accelerations are found in DECIGO. Moreover, the found accelerations peak at about $2$ orders of magnitude larger than missed accelerations, as well as the total detectable accelerations, which is consistent with the modest fraction of accelerations that are found.}
    \label{fig:losa_total}
\end{figure}

We show, in Figure~\ref{fig:losa_total}, the distribution of detectable (total), found, and missed accelerations. Of the total detectable BBHs in DECIGO, $12\%$ are found. The detectable accelerations follow a distribution that peaks between $10^{-17}\mathrm{s}^{-1}$ and $10^{-16}\mathrm{s}^{-1}$ with the median value being $1.7 \times 10^{-16} \mathrm{s}^{-1}$ and 90\% CI being $[4.7 \times 10^{-18}, \, 4.8 \times 10^{-15}]\mathrm{s}^{-1}$. 
The missed acceleration distribution peaks roughly at a similar value, having a median value: $8.5 \times 10^{-17} \mathrm{s}^{-1}$ and 90\% CI: $[2.5 \times 10^{-18},\, 1.8 \times 10^{-15}]\mathrm{s}^{-1}$ with relatively smaller support between $10^{-15}\mathrm{s}^{-1}$ and $10^{-14}\mathrm{s}^{-1}$.
Conversely, the found acceleration distribution peaks at $\sim 10^{-15}\mathrm{s}^{-1}$, having a median value: $6.3 \times 10^{-16} \mathrm{s}^{-1}$ and 90\% CI: $[3.5 \times 10^{-17},\, 1.3 \times 10^{-14}]\mathrm{s}^{-1}$.

\begin{figure}
    \centering
    \includegraphics{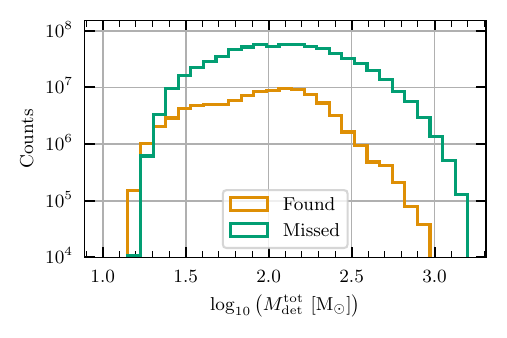}
    \caption{Weighted distributions of detector frame total mass ($M_\mathrm{det}$) of detectable BBHs with found and missed accelerations in DECIGO. The vertical axis shows counts in arbitrary units. Lighter BBHs provide stronger constraints on the accelerations (specifically, LOSAs), due to the increased number of cycles in-band. This is reflected in the distribution of found total masses having relatively less support at higher masses. The median of the found distribution is also smaller than the median of the missed distribution since the former is $\sim 61 \mathrm{M}_{\odot}$ while the latter is $\sim 108 \mathrm{M}_{\odot}$. 
    }
    \label{fig:mdet_total}
\end{figure}

We depict, in Figure~\ref{fig:mdet_total}, the distributions of found and missed detector-frame total masses $M_{\mathrm{det}} = M_{\mathrm{source}}(1 + z)$, where $z$ is the cosmological redshift. The distribution of found $M_{\mathrm{det}}$ is shifted towards smaller values relative to the corresponding missed distribution. This can be explained as follows. Lower-redshift mergers are lower-mass because of the mass-dependence of delay times and since mass segregation in GCs favors higher-mass mergers at early times \citep[e.g.,][]{Chatterjee2017a,FragioneRasio2023}. Smaller masses then enable better constraints on $a/c$ by virtue of spending more cycles in the detector band. 

\begin{figure}
    \centering
    \includegraphics{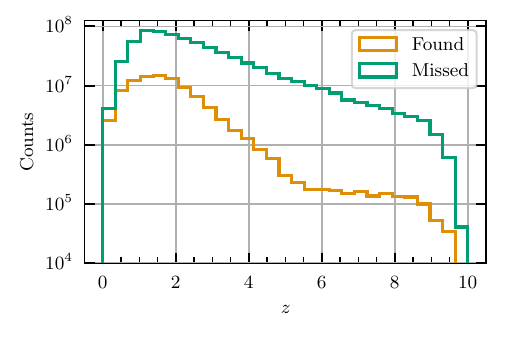}
    \caption{Weighted distributions ( vertical axis shows counts in arbitrary units) of redshifts of detectable BBHs with found and missed accelerations in DECIGO. The redshift distribution {of found BBHs} peaks at smaller values compared to the missed redshift distributions. This is because the r.m.s error on the acceleration estimate scales inversely with SNR, which in turn scales inversely with luminosity distance $d_\mathrm{L}$.}
    \label{fig:z_total}
\end{figure}

Similarly, Figure~\ref{fig:z_total} gives the distributions of found and missed BBH redshifts. Again, as with $M_{\mathrm{det}}$, the distribution of found $z$ is shifted towards smaller values relative to the corresponding missed distribution. Smaller $z$ correspond to larger SNRs, which reduces the r.m.s error approximately as $\Delta(a/c) \propto 1/\rho$. This allows relatively smaller accelerations to also be identified within $68\%$ confidence.

\subsection{Effect of GC properties on the distributions of found and missed acceleration}

\begin{figure*}
    \centering
    \includegraphics{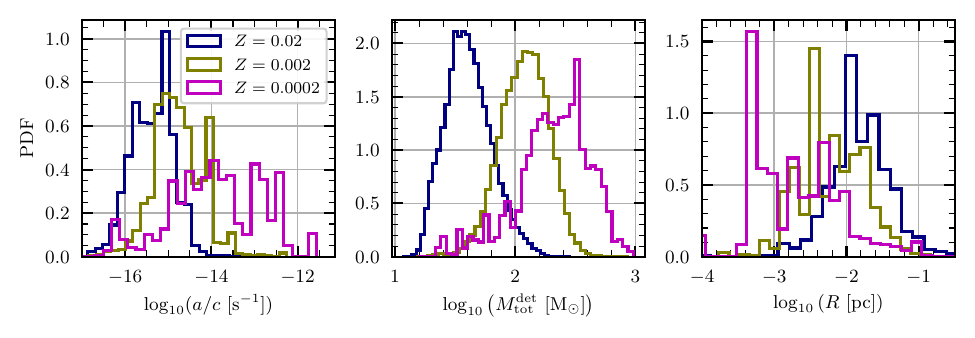}
    \caption{Weighted distributions of accelerations, detector frame masses $M_\mathrm{det}$, and outer orbital radii ($R$) of {the found} BBHs in GCs, with varying GC initial metallicity $Z$ in DECIGO. Larger $Z$ clusters have a relatively larger fraction of lighter BBHs. This is reflected in the fact that the majority ($\sim 93\%$) of found accelerations come from relatively higher $Z$ clusters (left panel). Moreover, the lighter masses enable probes of smaller accelerations due to the increased number of cycles in-band. Heavier masses require larger accelerations to be found. This is reflected in the shift of found accelerations towards the larger values with decreasing $Z$. It is further corroborated by the distributions of detector frame masses $M_\mathrm{det}$ (centre panel), where decreasing $Z$ pushes the distributions of found $M_\mathrm{det}$ to larger values. The distributions of outer orbital radii $R$ (right panel) are also consistent with the left and centre panels, where the distributions are peaking at systematically larger values with decreasing $Z$. Smaller $Z$ implies BBHs in such GCs need larger accelerations to be found, and therefore need to be closer to the centre of the potential.}
    \label{fig:losa_Z}
\end{figure*}

Properties of the GC determine the population of BBHs in the GC and its spatial distribution, and thus, by extension, the distribution of BBH accelerations. Here, we break down the effect of metallicity, virial radius, and galactocentric radius on the distribution of found and missed accelerations, and corresponding distributions of $M_{\mathrm{det}}$ and $R$ (outer orbital radius/cluster-centric radius).

The \textsc{cmc} catalog {encompasses} $3$ distinct GC metallicities: $Z = 2 \times 10^{-4}, 2 \times 10^{-3}, 2 \times 10^{-2}$. We extract BBH accelerations and construct distributions (weighted by the metallicity, cluster mass, and detectability weights) pertaining to found accelerations for each of these metallicities. We find that the majority of found accelerations, $93\%$, come from relatively higher metallicity GCs, $Z = 0.02 (29\%), 0.002 (64\%)$, with the fraction dropping to $7\%$ for $Z = 0.0002$.

The found distributions of accelerations, detector-frame masses, and orbital radii, are shown in Figure~\ref{fig:losa_Z}. The left panel shows a systematic preference for higher accelerations with decreasing metallicity. This can be understood from the fact that GCs with a larger metallicity prefer forming at low redshift and have a relatively larger fraction of low-mass BBHs. This decreases the detector-frame mass and enables measurements of smaller accelerations.
The larger detector-frame mass BBHs in low-metallicity (and high-redshift) clusters, need larger accelerations to be confidently identified (found) as accelerating.
This explanation is further corroborated by the corresponding distributions in the center panel, which show a systematic shift of $M_{\mathrm{det}}$ distributions to larger values with decreasing metallicity---the medians being $\sim 34 \mathrm{M}_{\odot}$, $\sim 99 \mathrm{M}_{\odot}$, and $\sim 125 \mathrm{M}_{\odot}$ in descending order of the metallicity. The right panel is also consistent with this picture since the distribution of outer orbital radii shifts to decreasing values with decreasing metallicity\footnote{See Appendix~\ref{app:weights} for more details.}. Smaller radii yield larger accelerations, which are required by heavier masses to be identified confidently (found) as accelerating. 

We study the effect of changing the virial radius on the distributions of found accelerations and corresponding $M_{\mathrm{det}}$ and $R$. The \textsc{cmc} catalog provides $4$ discrete values: $r_v/\mathrm{pc} = 0.5, 1.0, 2.0, 4.0$. The effect of changing $r_v$ on the distributions is less pronounced than the effect of changing $Z$. This can be explained as follows. For a given mass distribution and location of BBH mergers in a GC, a smaller $r_v$ leads to more compact GCs, which in turn leads to larger accelerations. However, while there is a direct correlation between $r_v$ and acceleration, there is no such correlation between $r_v$ and BBH mass. Thus, while smaller $r_v$ yield larger accelerations in general, they do not necessarily yield smaller $M_{\mathrm{det}}$ which enables stronger constraints on $a/c$. Nevertheless, we do find that the fraction of found accelerations, varies markedly with decreasing $r_v$: $r_v = 0.5 \mathrm{pc}~(70\%), r_v = 1.0 \mathrm{pc}~(23\%), r_v = 2.0 \mathrm{pc}~(6\%), r_v = 4.0 \mathrm{pc}~(1\%)$.

\begin{figure*}
    \centering
    \includegraphics{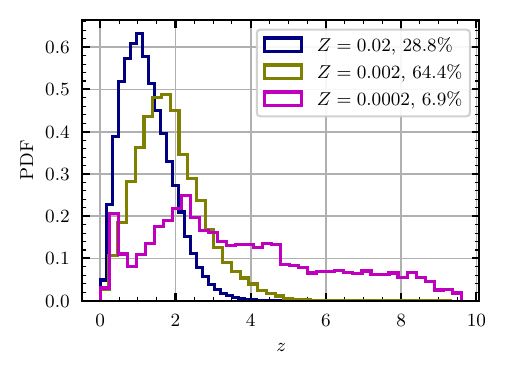}
    \includegraphics{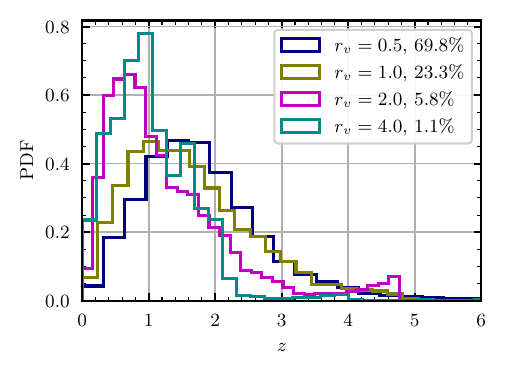}
    \caption{Weighted distributions of {redshifts ($z$s) of found BBHs} in DECIGO as a function of metallicity $Z$ (left panel) and virial radius $r_v$ (right panel). GCs with larger $Z$ are younger (i.e., they are formed at lower $z$). Moreover, they have a larger fraction of lighter BBHs, which are only detectable at lower $z$s. On the other hand, smaller $Z$ clusters are older and have a relatively larger fraction of heavier BBHs, which are detectable at higher $z$. Both of these contribute to redshift distributions in low $Z$ GCs having support at large $z$. On the other hand, GCs with smaller $r_v$ have BBHs with relatively larger accelerations, which enables them to be found at larger $z$. This explains the systematic shift of the redshift distribution of found accelerations towards larger values with decreasing $r_v$.}
    \label{fig:z_rv_Z}
\end{figure*}

We additionally study the effect of varying $Z$ and $r_v$ on the $z$ distribution pertaining to found accelerations. This is shown in Figure~\ref{fig:z_rv_Z}. The left panel shows the $z$ distribution getting progressively larger support at larger $z$ values with decreasing metallicity $Z$. This can be readily explained in terms of the age of clusters. Lower metallicity GCs are older and thus reside at larger $z$ values. Conversely, higher metallicity GCs are younger and contain a larger fraction of lower-mass BBHs. This results in fewer samples of found accelerations at larger $z$ -- both due to reduced SNR as well as poorer acceleration constraints from higher-mass BBHs. The effect of $r_v$ on $z$ distributions is less pronounced ({right panel of Figure} \ref{fig:z_rv_Z}), although larger accelerations from smaller {values of} $r_v$ imply increasing support at larger redshifts. 

We do not find any significant effect of changing $r_g$ on the distributions or the fraction of found accelerations. This is due to the fact that the accelerations extracted from the \textsc{cmc} simulations consider only the potential of the GC and not the potential of the galaxy in which the GC is hosted. The center of mass of the GC itself will have an acceleration, which depends on $r_g$ but has not been considered in this work. Adding this effect will likely cause a systematic shift in the acceleration distributions; however, we do not expect the distributions to be impacted significantly if the GC is situated at typical locations ($r_g \sim {\rm kpc}$) in the galaxy\footnote{See Figure 5 of \citealt{vijaykumar2023} for an estimate of acceleration due to the gravitational potential of a Milky Way-like galaxy.}.

We refer the reader to Appendix~\ref{app:weights} for a more detailed explanation of how the application of metallicity and cluster-mass weights to the intrinsic distribution of found accelerations impact the variation of the fraction of BBHs with these accelerations as a function of cluster properties. 

\begin{figure*}
    \centering
    \includegraphics{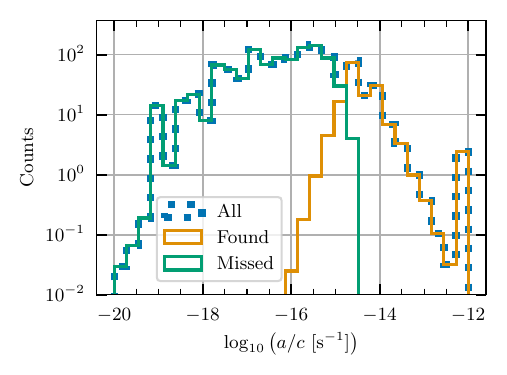}
    \includegraphics{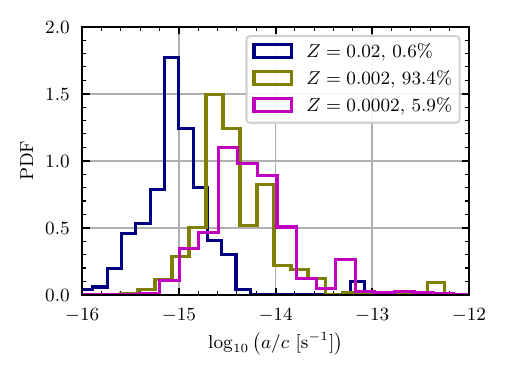}
    \caption{Weighted distributions of detectable found and missed accelerations in LISA (left panel; vertical axis shows counts in arbitrary units), as well as the effect of metallicity $Z$ on the found distributions (right panel). The fraction of found accelerations is $\sim 14\%$, and the acceleration distributions span similar values as for DECIGO (see Figure ~\ref{fig:losa_total}). The imprint of $Z$ on the distribution of found accelerations is different from what was found for DECIGO (see Figure~\ref{fig:losa_Z}). $Z = 0.002$ dominates the fraction of measurable accelerations. This can be explained as the result of competing effects. Higher $Z$ has a larger fraction of lighter BBHs, many of which are not detectable in LISA. On the other hand, lighter BBHs enable better constraints on acceleration. Among the discrete metallicities considered, the one closest to the ``sweet spot'' that has detectable BBHs with measurable accelerations is $Z = 0.002$.}
    \label{fig:losa_total_LISA}
\end{figure*}

\subsection{Distributions of found and missed accelerations in LISA}

LISA's sensitivity band covers a frequency range that is lower than DECIGO: $f \in [10^{-4}, 1] \mathrm{Hz}$. The BBHs, therefore, spend a significantly longer time within the LISA band than the DECIGO band, which should enable stronger constraints on acceleration. However, LISA's sensitivity to stellar mass BBHs is much lower as compared to BBHs. As a result, the majority of the lighter BBHs are not detectable ($\rho < 8$) in LISA, given that the Madau-Fragos SFRD peaks at $z\sim 2$. Nevertheless, among those BBHs that are detectable, $\sim 14\%$ are found, in part because the lower frequency reach of LISA enables binaries to spend longer times in-band. 

In Figure~\ref{fig:losa_total_LISA}, we provide the distribution of found and missed accelerations (left panel), and the variation of found acceleration distributions with metallicity (right panel), whose imprint was found to be the most pronounced in the DECIGO analysis. We once again see that the found acceleration's distribution peaks between $10^{-15}\mathrm{s}^{-1}$ and $10^{-14}\mathrm{s}^{-1}$ with the median value being $1.2 \times 10^{-15} \mathrm{s}^{-1}$ and 90\% CI being $[3.6 \times 10^{-17}, \, 1.7 \times 10^{-14}]\mathrm{s}^{-1}$. However, unlike DECIGO, we find that the majority of these accelerations ($93\%$) come from $Z = 0.002$ clusters. This can be explained as the consequence of competing effects. {GCs with} higher metallicities have a larger fraction of lighter BBHs, many of which are undetectable with LISA. On the other hand, {BBHs with} lighter masses enable more precise acceleration measurements. Among the discrete metallicities considered, the metallicity value closest to the ``sweet spot'' that has both detectable and measurable accelerations is $Z = 0.002$. It should be noted that the metallicity weight (and to a lesser extent the cluster-mass weight) also contributes to enhancing the fraction of found accelerations for $Z = 0.002$ (see Appendix~\ref{app:weights}). Correlations of metallicity with $M_{\rm det}$ and $R$ are similar though less pronounced than what was found for DECIGO, and are therefore not plotted. 

{Given that the sensitivities of other proposed millihertz space-based detectors such as TianQin \citep{Luo_2016} are similar to LISA, we do not expect our forecasts to differ significantly for those detectors.}
 
\section{Summary and Outlook}\label{sec:conclusions}

GCs  are one class of dense stellar environments expected to host BBH mergers. The $\sim 90$ events detected by the LIGO-Virgo-KAGRA network cannot conclusively determine if a given BBH was hosted by a GC, although the merger rate in GCs can be estimated by comparing against GC simulations \citep{2021RNAAS...5...19R}, or by calculating the fraction of the BBH population that is consistent with having isotropic spin directions \citep{Fishbach:2023xws}.
Such rates, however, are limited by the sample size of the detected BBHs, as well as uncertainties in the models of GCs and their initial properties (size, metallicity, etc).

On the other hand, LOSAs of BBHs leave an imprint on their GW waveform at -4PN, and can therefore be potentially constrained by detectors sensitive at low frequencies (e.g: decihertz, millihertz bands) such as DECIGO and LISA. BBHs in GCs are expected to contain finite LOSAs, and their distribution could contain imprints of the properties of the GCs. LOSAs could therefore assist in identifying the provenance of BBHs.

\begin{figure}
     \centering
     \includegraphics{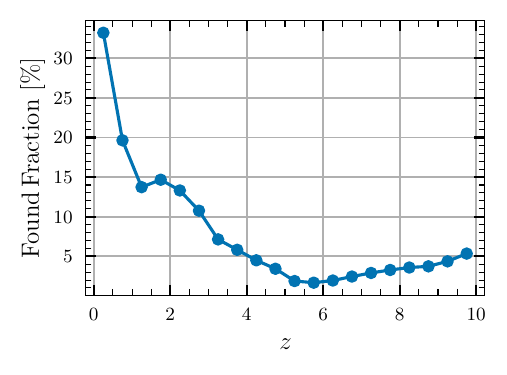}
     \caption{Variation of the fraction of found accelerations with redshift in DECIGO. Each fraction corresponds is calculated in a redshift bin of width 0.5, with each point sitting at the centre of that bin. The found fraction initially decreases, reaches its minimum value in the redshift bin [5.5, 6], and starts slightly rising again due to increasing metallicity weight. }
     \label{fig:frac_var_z}
 \end{figure}

In this work, we forecast the distribution of detectable BBHs in GCs in DECIGO and LISA eras, that also produce accelerations that are identifiable (found) at $\geq 68\%$ confidence. To do so, we use the outputs of the \textsc{cmc} catalogue to extract distributions of BBH accelerations, following {the scheme presented} in Figure~\ref{fig:flowchart}. We summarize our main results below.
\begin{enumerate}
 \item We find that $\sim 12\%$ ($\sim 14\%$) of detectable BBHs in the DECIGO (LISA) era have accelerations that are well-constrained away from zero. We also find that the distribution of measurable (found) accelerations peaks at $10^{-15}\mathrm{s}^{-1}$ in DECIGO and between $10^{-15}\mathrm{s}^{-1}$ and $10^{-14}\mathrm{s}^{-1}$ in LISA.

  \item Among found accelerations, the majority ($\sim 93\%$ in DECIGO and LISA) come from relatively higher metallicity ($Z = 2 \times 10^{-2}, 2 \times 10^{-3} $) clusters. This is clearly reflected in the mass spectrum of BBHs with found accelerations. Higher metallicity clusters form at low redshift and have a larger fraction of relatively low-mass BBHs, thus enabling better measurements of acceleration. Conversely, low metallicity ($Z = 2 \times 10^{-4}$) results in a larger fraction of high (detector frame) mass BBHs, and their accelerations need to be $1-2$ orders of magnitude larger to be found. In LISA, $Z = 0.002$ dominates the fraction of measurable accelerations due to competing effects of lighter masses being more difficult to detect while also enabling more precise acceleration measurements. 
 
 \item We observe correlations between the virial radius $r_v$ of the cluster and the shape of the distributions, although these are less pronounced compared to the correlations with metallicity. Nevertheless, the majority of the found accelerations come from small $r_v$ (e.g. $70\%$ of found accelerations come from $r_v = 0.5 \mathrm{pc}$). We find no appreciable dependence of the fraction of identifiable accelerations on the galactocentric radius $r_g$, likely because the accelerations extracted from the \textsc{cmc} simulations do not account for the galactic potential that hosts the GC.
\item Converting the percentage of found accelerations to a rate of found accelerations in the DECIGO/LISA eras requires estimates of BBH merger rates out to redshifts $z > 1$, which to date is poorly constrained. We instead plot the fraction of found accelerations in DECIGO\footnote{Given that the intrinsic rate of detectable stellar mass BBHs is expected to be small in the LISA era, we do not plot the corresponding evolution of the fraction of found accelerations with $z$. All events with found acceleration in LISA lie at $z \lesssim 0.2$.} as a function of redshift in Figure~\ref{fig:frac_var_z}. This fraction initially decreases, reaches its minimum value in the redshift bin [5.5, 6], and starts rising slightly again. This rise coincides with the redshift ($z\sim 6$) at which $Z=0.0002$ clusters overtake $Z=0.002$ clusters in their contribution to the total number of detected events (in part due to high $W_Z$; see Figure \ref{fig:metallicity-weight}). Since the source-frame masses in $Z=0.0002$ clusters are slightly higher than those in $Z=0.002$ clusters, and events in low-metallicity clusters have higher acceleration owing to their relative closeness to the center\footnote{This is due to lower natal kicks in low-metallicity environments \citep{cmc-paper}. See also Appendix \ref{subsec:W_cl} for a related discussion.}, the number of found events increases slightly in comparison to the number of missed events above $z\sim 6$\footnote{ The slight rise around $z\sim 1.5$ can be similarly attributed to the redshift beyond which binaries in $Z=0.002$ clusters dominate over those in $Z=0.02$ clusters. }.
\end{enumerate}
We note that the results mentioned above and in the rest of the work are contingent on our modeling assumptions for $W_{Z}$, $W_{\rm cl}$, and $W_{\rm det}$. For instance, our understanding of cosmic GC formation history is incomplete, and the assumption that GC formation follows star formation might not be a good one. Semi-analytic models of GC formation built using dark matter halo merger trees \citep{2019MNRAS.482.4528E} show that the cluster formation rate density peaks at a higher redshift ($z \sim 4$) and does not track the SFRD. However, these estimates are themselves model-dependent, and we prefer to use an observation-oriented (ie. the Madau-Fragos SFRD) prescription in our work. While we only focus on model-dependent forecasts of  LOSAs, measurements of LOSAs can also be used to constrain host GC properties of BBHs independently or in tandem with methods in \citealt{Fishbach:2023xws}.
Binaries in GCs, especially those that merge in the cluster cores, are expected to have non-negligible orbital eccentricities. Although we do not consider the effect of orbital eccentricity in our SNR calculations or Fisher forecasts, the number of mergers with found accelerations could increase if we include these effects~\citep{Xuan:2022qkw}.

Other dense stellar environments that could host BBHs include nuclear star clusters \citep{hoang2018} and AGNs \citep{ford2022}. As follow-up work, we plan to study the distributions of accelerations of BBHs in these environments, and the imprints of their properties on said distributions. We also plan to compare distributions of accelerations coming from these different dense stellar environments, which, in principle, could help in determining the provenance of the BBHs. 

The accelerations of BBHs extracted from the \textsc{cmc} simulations consider only the effect of the GC gravitational potential. However, encounters of BBHs with a third body, when they lie within the band of the detectors, could impart an acceleration that is significantly larger than those provided by the GC potential. Accelerations of such in-encounter mergers could therefore be detectable even by future ground-based detectors, such as the XG network. We plan to investigate this as well in future work.

\section*{Acknowledgements}

We thank Michael Zevin for comments on a draft version of this work. We also thank Zoheyr Doctor, Nathan Johnson-McDaniel, Parameswaran Ajith, and K. G. Arun for useful discussions. This work has made use of \texttt{NumPy} \citep{vanderWalt:2011bqk}, \texttt{SciPy} \cite{Virtanen2019}, \texttt{astropy} \citep{2013A&A...558A..33A, 2018AJ....156..123A}, \texttt{Matplotlib} \citep{Hunter:2007},
\texttt{seaborn} \citep{Waskom2021}, \texttt{jupyter} \citep{jupyter}, and \texttt{pandas} \citep{mckinney-proc-scipy-2010} software.  A.V. is supported by the Department of Atomic Energy, Government of India, under Project No. RTI4001.  G.F. acknowledges support by NASA Grant 80NSSC21K1722 and NSF Grant AST-2108624 at Northwestern University. SC acknowledges support from the Department of Atomic Energy, Government of India, under project no. 12-R\&D-TFR-5.02-0200 and RTI 4002. All computations were performed on the SARATHI computing cluster at IUCAA.

\section*{Data Availability}

The data underlying this article will be shared on reasonable request to the corresponding author.



\bibliographystyle{mnras}
\bibliography{acceleration} 




 \appendix
 \section{Effect of weights on the distribution of measurable accelerations}\label{app:weights}
\begin{figure*}
    \centering    \includegraphics{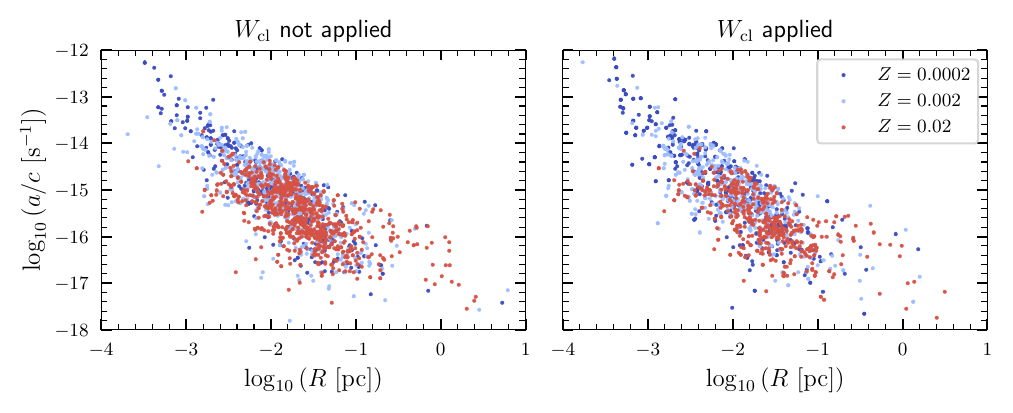}
    \caption{Scatter plots of detectable accelerations vs corresponding distance of the merger from the center of the cluster, colored by metallicities. The left panel shows mergers in clusters of different metallicities when 2000 mergers (with detectable accelerations) are drawn randomly without any weights, while the right panel shows the same when mergers are drawn following the initial cluster mass weight $W_{\rm cl}$.
    Notably, as explained in section \ref{subsec:W_cl}, binaries in high metallicity clusters are farther away from the center on average and the number of binaries detected in high metallicity environments decreases when the initial cluster mass weight is applied. In particular, $Z = 0.02$ binaries make up $46\%$ of the total number in the left panel as opposed to $27\%$ in the right panel. }
    \label{fig:cluster-noweights-weights}
\end{figure*}

 \subsection{Initial Cluster Mass Weight}
 \label{subsec:W_cl}
To construct the distribution and determine the fraction of found (measurable) accelerations, we apply a weight $W_{\mathrm{cl}}$ that is inversely proportional to the square of the initial cluster mass (cf. Eq.~\ref{eq:cluster-weight}). The effect of applying this weight is to enhance the fraction of found accelerations pertaining to low metallicity mergers. This can be explained as follows:
\begin{enumerate}
    \item High metallicity environments form low mass pre-supernova cores due to higher line-driven winds~\citep{Vink:2001cg}.
    \item Low mass cores get a larger supernova natal kick owing to lesser mass fallback~\citep{fryer2012}. This high kick displaces them from the center of the cluster, \textit{i.e.} to higher $R$, possibly also ejecting them from the cluster in the process.
    \item The only way to then have appreciable acceleration for high metallicity mergers is by having a very dense environment, i.e. clusters with a higher mass.
    \item Since massive clusters are down-weighted by $W_{\rm cl} \sim 1/M^2_{\rm cl}$, the total number of high-metallicity mergers is also down-weighted.
\end{enumerate}

This is illustrated in the scatter plots of found BBHs in DECIGO. Figure~\ref{fig:cluster-noweights-weights} shows scatter plots of found accelerations vs corresponding radii for different metallicities, with and without accounting for $W_{\rm cl}$. 

\begin{figure}
    \centering
    \includegraphics{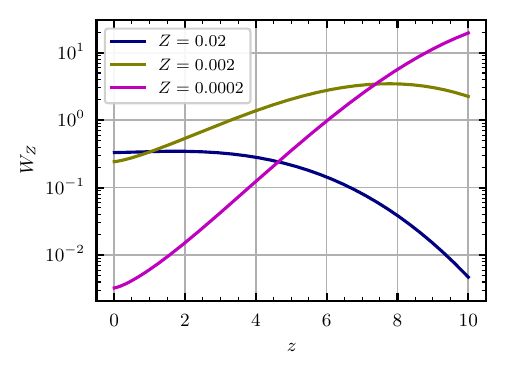}
    \caption{Metallicity weights pertaining to found accelerations as a function of cluster-formation redshift ($z$), for three discrete values of $Z$. The effect of these weights is broadly a convolution of two distributions -- A lognormal distribution whose mean is a function of redshift, and the Madau-Fragos SFRD from which the redshift distributions are drawn. The case $Z = 0.002$ has the largest weight values for the redshift regime where the Madau-Fragos SFRD has maximum support $z \in [1, 4]$. This in part explains why $Z = 0.002$ has the largest fraction of found accelerations.}
    \label{fig:metallicity-weight}
\end{figure}

\subsection{Metallicity Weight}

Another weight that is applied to the distribution of found accelerations is the metallicity weight. The weight is evaluated using a log-normal distribution in the metallicity whose mean is redshift dependent \citep{madaufragos2017}. Since the BBH redshifts are drawn following the Madau-Fragos SFRD, the metallicity weight is (broadly) a result of convolving this distribution with the log-normal distribution.  

We plot metallicity weights for found samples as a function of redshift. We see that $Z = 0.002$ has the largest weights between $z = 1 - 4$, in comparison to the other metallicities $Z = 0.02$ and $ 0.002$. Since the Madau-Fragos SFRD has the largest support between $z = 1 - 4$, metallicity weights tend to enhance the fraction of found accelerations for low metallicities (say $Z = 0.002$), relative to the other metallicities. This in part explains the fractions displayed in Figure~\ref{fig:losa_Z}. Furthermore, $Z = 0.0002$ has the largest weights only at $z \gtrsim 7.5$, where the Madau-Fragos SFRD has negligible support. On the other hand, where the SFRD has the largest support ($z = 1 - 4$), this metallicity value has the smallest weight. This explains, in part, the small fraction of found accelerations assigned to $Z = 0.0002$.

\subsection{Impact of randomized orbital inclination and stringent measurability criteria}\label{app:random_inclination}

\begin{figure*}
    \centering
    \includegraphics{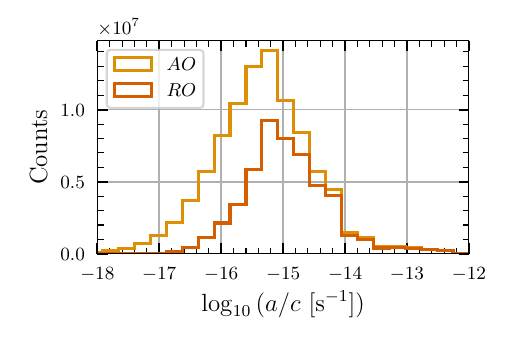}
    \includegraphics{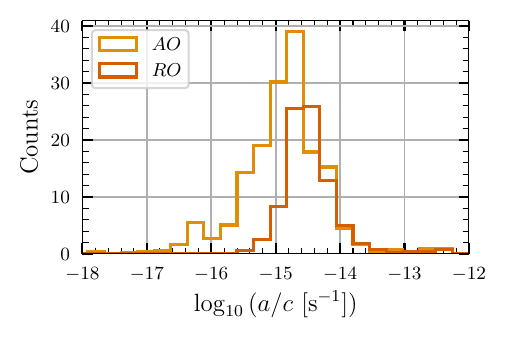}
    \caption{{The left panel shows a comparison between the distributions of found accelerations (absolute values) for the aligned orientation (AO) and random orientation (RO) cases in DECIGO, while the right panel shows the same in LISA. We observe that the found fraction drops to $\sim 6\%$ and $\sim 7\%$ in DECIGO and LISA, respectively.}}
    \label{fig:losa_rand_orient}
\end{figure*}

\begin{figure*}
    \centering
    \includegraphics{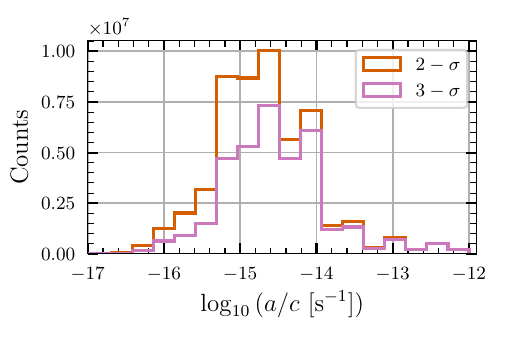}
    \includegraphics{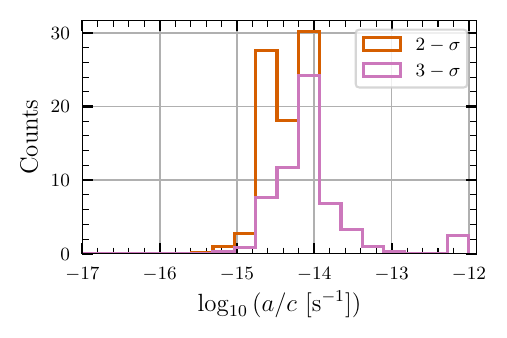}
    \caption{{The left panel shows the distributions of found accelerations for the $2-\sigma$ and $3-\sigma$ detections in DECIGO, while the right panel shows the same in LISA. We again observe the drops in the found fractions in both detectors.}}
    \label{fig:losa_2-3_sig_comb}
\end{figure*}

{The analysis presented in this paper assumes the acceleration vector to be aligned with the line of sight. The resulting fractions of found accelerations should therefore be thought of as upper limits. To assess the reduction in this fraction from a more realistic set-up, we allow the angle $\theta$ with respect to the line of sight to vary uniformly in $\cos\theta$. We find that the fraction of measurable accelerations reduces to $\sim 6\%$ in DECIGO and $\sim 7 \%$ in LISA (see Figure \ref{fig:losa_rand_orient}). We also present results for accelerations measurable at $2-\sigma$ and $3-\sigma$ confidence (see Figure \ref{fig:losa_2-3_sig_comb}). We find that the fraction of found accelerations drops to $\sim 7 \%$ and $\sim 5\%$ for the $2-\sigma$ and $3-\sigma$ detections, respectively, in DECIGO while the same fractions in LISA drop to $\sim 8 \%$ and $\sim 5 \%$.}

\bsp	
\label{lastpage}
\end{document}